 \documentclass[twocolumn,showpacs,preprintnumbers,amsmath,amssymb,showkeys,superscriptaddress]{revtex4}
 \usepackage{graphicx}
 \usepackage{dcolumn}
 \usepackage{bm}
\def\bd{\begin{document}} \def\ed{\end{document}}
\def\bw{\begin{widetext}}  \def\ew{\end{widetext}}
\def\beq{\begin{equation}} \def\eeq{\end{equation}}
\def\bea{\begin{eqnarray}} \def\eea{\end{eqnarray}} \def\rt{{\rm t}}
\def\bcc{\begin{center}} \def\ecc{\end{center}}
\def\f{\left} \def\g{\right} \def\e{{\rm e}}
\def\p{\partial} \def\cov{{\rm cov}} \def\ch{{\rm ch}}
\def\vf{\varphi} \def\EE{e$^+$e$^-$}  \def\pt{p_{\rm t}}
\def\cl{\centerline} \def\fnsz{\footnotesize}
\def\ub{\underbar} \def\hs{\hskip} \def\vs{\vskip} \def\ni{\noindent}
\def\pa{\parindent} \def\ej{\vfill\eject} \def\hf{\hfill}
\def\la{\langle} \def\ra{\rangle} \def\r#1{$^{[#1]}$}
\def\qqg{q$\bar{\rm q}$g\hskip1pt} \def\epn{e$^+$e$^-$\hskip1pt }
\def\pt{{p_{\rm t}}} \def\vf{\varphi} \def\yct{y_{\rm cut}}
\def\NAtwo{{\sc na}{\footnotesize 22}}  \def\QGP{{\sc qgp}}
\def\NPM{{\sc npm}} \def\NFM{{\sc nfm}} \def\EFM{{\sc efm}}
\def\RHIC{{\sc rhic}}  \def\BNL{{\sc bnl}}  \def\SF{{\sc sf}} \def\DF{{\sc df}}
\def\d{{\rm d}} \def\ebe{event-by-event} \def\JACEE{{\sc jacee}}
\bd

 \title{A Comparison of different measures for dynamical event-mean \\
 transverse momentum fluctuation}
 \thanks{This work is supported in part by the NSFC under project
90103019.}

\author{Liu Lianshou}

\affiliation{Institute of Particle Physics, Huazhong Normal
University, Wuhan 430079 China}

\email{liuls@iopp.ccnu.edu.cn}

\author{Fu Jinghua}

\affiliation{Engineering Physics Department, Tsinghua University,
Beijing 100084, China}

\affiliation{Institute of Particle Physics, Huazhong Normal
University, Wuhan 430079 China}

\begin{abstract}
Various measures for the dynamical event-mean transverse momentum
fluctuation are compared with the real dynamical fluctuation using
a Monte Carlo model. The variance calculated from the $G$-moments
can reproduce the dynamical variance well, while those obtained
through subtraction procedures are approximate measures for not
very low multiplicity. The $\Phi_\pt$ can also serve as an
approximate measure after divided by the square root of mean
multiplicity.
\end{abstract}



\maketitle

The successful operation of the relativistic heavy ion collider
\RHIC\ at \BNL\ has stimulated an intensive study on the
theoretically predicted quark gluon plasma (\QGP) phase
transition. It is expected that the appearance of phase transition
will result in an anomalous behavior in the \ebe\ fluctuations of
transverse momentum~\cite{Mrowski}\cite{Shuryak}\cite{Pisarski}.

Since the multiplicity in a single event is finite, there are
inevitably statistical fluctuations (\SF) in the \ebe\ analysis of
transverse momentum. Various
methods~\cite{PHENIX}\cite{Voloshin}\cite{CERES} have been
proposed to get rid of the statistical fluctuations and measure
the dynamical ones. Most of them are based on a {\it subtraction}
procedure, {\it i.e.} first to estimate the variance of
statistical fluctuations and then subtract it from that of the
experimental data. The basic problem in this kind of methods is
how to reliably estimate the statistical variance.

Another approach proposed recently~\cite{FuLiu} is to {\it
eliminate} the \SF\ directly from the experimental data. This
method can reproduce the dynamical variance exactly provided the
\SF\ are Poissonian, {\it i.e.} the particles are emitted
independently.

The aim of this paper is to compare the different measures
proposed in the market and, in particular, to examine the
reliability of the estimation of statistical variance. The so
called $\Phi_{\pt}$ measure~\cite{marek} will also be discussed
and a new measure $\Phi^{(\rm r)}_{\pt}$ will be proposed, which
can largely reduce the strong multiplicity dependence of
$\Phi_{\pt}$ itself.

Let $p(\pt)$ be the dynamical transverse-momentum distribution in
single events and  
\beq p_m=\int_{\delta_m} p(\pt) \d \pt, \qquad (m=1,2,\dots, M)  \eeq 
be the corresponding ``coarse-grained'' distribution~\cite{BP}.
The realization of $p(\pt)$ in experiment is the distribution of
the total number $n$ of particles in the $\pt$ region $\Delta$,
and the expression 
\beq q_m = {n_m}/{n} \qquad (m=1,2,\dots, M) \eeq 
is an evaluation of $p_m$, where $n_m$ is the number of particles
falling into the $m$th bin. Thus the dynamical and
experimentally-measured event-mean transverse momentum are,
respectively, 
\beq \bar {\pt}_{\rm dyn} = \int_\Delta \pt p(\pt) \d \pt =
\sum_{m=1}^M
(\pt)_m p_m, \eeq 
\beq \bar {\pt}_{\rm exp}  = \sum_{m=1}^M (\pt)_m q_m=\sum_{m=1}^M
(\pt)_m
\frac{n_m}{n}, \eeq 
where $(\pt)_m$ is the $\pt$ value in the $m$th bin. The
event-space moments of $\bar {\pt}_{\rm dyn}$ and $\bar {\pt}_{\rm exp} $ are 
\beq C^{\rm dyn}_p(\bar {\pt}) = \la \bar{\pt}_{\rm dyn} ^p\ra =
\f\la \f(\sum_{m=1}^M (\pt)_mp_m\g)^p \g\ra,
\eeq 
\beq C^{\rm exp}_p(\bar{\pt}) = \la (\bar{\pt}_{\rm exp})^p\ra
= \f\la \f(\sum_{m=1}^M (\pt)_m\frac{n_m}{n}\g)^p  \g\ra, \eeq 
respectively. These two are evidently unequal due to the existence
of \SF. In particular, the experimental moments $C_p^{\rm
exp}(\bar{\pt})$ depend crucially on the particle
multiplicity~\cite{NA49}, which is mainly due to \SF. Therefore,
subtraction procedures are commonly used to extract the dynamical
fluctuation of $\bar{\pt}$.

The basic idea of the subtraction procedure is that the variances
$\sigma^2(\bar\pt)=C_2(\bar\pt)-C_1(\bar\pt)^2$
of dynamical and statistical fluctuations are additive 
\beq  \sigma^2_{\bar{\pt}{\rm data}} =\sigma^2_{\bar{\pt}{\rm
dyn}} +\sigma^2_{\bar{\pt}{\rm stat}}, \eeq 
where $\sigma^2_{\bar{\pt}{\rm data}}$, $\sigma^2_{\bar{\pt}{\rm
dyn}}$ and $\sigma^2_{\bar{\pt}{\rm stat}}$ are the experimental,
dynamical and statistical variances, respectively. This additivity
is true if the \DF\ and \SF\ are independent, which is a
reasonable assumption. Therefore, the main problem in the
subtraction procedure for extracting the dynamical variance is to
estimate the statistical ones. Various methods have been proposed
for this purpose.

For example, in Ref.~\cite{PHENIX} the results from mixed events
are considered as the baseline for the random distribution and the
difference in the fluctuation from a random distribution defined
as 
\beq d=\omega_{\rm data}-\omega_{\rm baseline} \eeq  
is taken as a measure of the dynamical fluctuation. In Eq.(8) 
\beq \omega=\frac{\sqrt{\la \bar{\pt}^2\ra-\la \bar{\pt}\ra^2}}
{\la\bar{\pt}\ra}=\frac{\sqrt{\sigma^2_{\bar{\pt}}}}{\la\bar{\pt}\ra}, \eeq 
$\bar{\pt}$ is the mean transverse momentum in a single event and
$\la\cdots\ra$ denotes the average over event sample.

Alternatively, in Ref.~\cite{Voloshin} the statistical variance of
event mean $\pt$, under the assumption of independent particle
production, is estimated as 
\beq \sigma^2_{\bar{\pt}{\rm stat}}=\frac{\sigma^2_{\pt{\rm
incl}}}{\la n\ra}, \eeq 
and, therefore, the dynamical variance is equal to
\beq \sigma^2_{\bar{\pt}_{\rm VKR}}=\sigma^2_{\bar{\pt}{\rm
data}}-\sigma^2_{\bar{\pt}{\rm stat}}=\sigma^2_{\bar{\pt}{\rm
data}}-\frac{\sigma^2_{\pt{\rm incl}}}{\la n\ra}. \eeq 
A somewhat different expression is used in Ref.~\cite{CERES},
where the experimental variance is weighted by event multiplicity
$N_j$: \beq \sigma^2_{\bar{\pt}_{\rm CERES}}=
\frac{\sum_{j=1}^{\cal N} N_j(\bar{\pt}^j-\la
\bar{\pt}\ra)^2}{\sum_{j=1}^{\cal N} N_j}
-\frac{\sigma^2_{\pt{\rm incl}}}{\la n\ra}, \eeq 
where $\cal N$ is the total number of events.

A widely used measure for the non-statistical mean $\pt$
fluctuation is the $\Phi_{\pt}$ proposed in Ref.~\cite{marek} 
\beq\Phi_{\pt} \equiv \sqrt{\la Z^2\ra/\la n \ra}
-\sqrt{\la{z^2}\ra}, \eeq 
where $z$ and $Z$ are defined as $z\equiv \pt-\la \pt\ra$ for each
particle and $Z\equiv \sum_{i=1}^n z_i = n(\bar{\pt}-\la\pt\ra)$
for each event, respectively~\cite{note2}. The second term in the
r.h.s. of Eq. (13) is the square root of the inclusive variance
$\sigma^2_{\pt{\rm incl}}=\la(\pt-\la\pt\ra)^2\ra$. Assuming that
the multiplicity fluctuation is uncorrelated with the $\pt$
fluctuation, we get from Eq.(13) 
\beq \Phi_{\pt}=\sqrt{\la n^2\ra/\la n \ra} \sigma_{\bar{\pt}{\rm
data}}-\sigma_{\pt{\rm incl}}. \eeq 
This equation is evidently similar to Eq.(11).

All of the above measures have the same structure, being based on
a subtraction procedure, {\it i.e.} to subtract the variance of
$\bar{\pt}$ or a quantity related to it, that will be expected
from a pure statistical system, from the same quantity obtained in
experiment. These measures will, of course, vanish for a pure
statistical system, and a non-vanishing value of them will
indicate the existence of dynamical effect. Therefore, the
measures based on the subtraction procedure, as those listed
above, will at least qualitatively measures the effect of \DF.

 \begin{figure}
 \includegraphics[width=7.2cm]{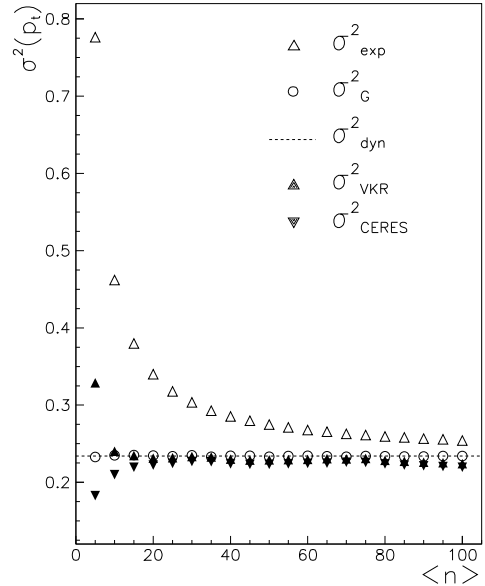}
 \caption{\label{fig:cgcp} The experimental (open triangles) and dynamical
 (dashed line) variances of $\bar\pt$, together with the variances calculated
 from $G$-moments (open circles) and from Eqs. (11) (upward solid triangles)
 and (12) (downward solid triangles).}
 \end{figure}

In order to have an idea on how the various subtraction methods
work quantitatively, let us carry on a Monte Carlo simulation
using a toy model. In this model the distribution of $\pt$ is taken as 
\beq p(\pt) = \frac{4}{a^2} \pt \e^{-2\pt/a}, \eeq 
with a Gaussian distributed parameter $a$ ($\sigma^2(a)=0.24$). In
total $20\times 500,000$ events are generated for 20 average
multiplicities. The resulting dynamical and experimental variances
$\sigma^2_{\bar{\pt}{\rm dyn}}$ and $\sigma^2_{\bar{\pt}{\rm
exp}}$ are plotted in Fig.1 as dashed line and open triangles,
respectively. For comparison, the $\sigma^2_{\bar{\pt}{\rm VKR}}$
and $\sigma^2_{\bar{\pt}{\rm CERES}}$ are also plotted in the same
figure as upward and downward solid triangles, respectively.

The differences between the experimental and dynamical variances
$\sigma^2_{\bar{\pt}{\rm exp}} - \sigma^2_{\bar{\pt}{\rm dyn}}$
are due to \SF. It can be seen from the figure that the
$\sigma^2_{\bar{\pt}{\rm VKR}}$ indeed reduces the \SF, especially
when the average multiplicity is not very low; while the
$\sigma^2_{\bar{\pt}{\rm CERES}}$ overshoots the influence of \SF.

On the other hand, the open circles shown in  the figure reproduce
the dynamical variances very well. They are obtained through a
totally different approach~\cite{FuLiu}, which is based on the
elimination of the \SF\ directly from the experimental data.

For the first order moment the elimination of \SF\ is
straightforward, 
\beq C^{(\rm dyn)}_1(\bar {\pt})=G_1(\bar {\pt}) = \f\la
\sum_{m=1}^M (\pt)_m \frac{n_m}{\la
n\ra} \g\ra,\eeq 
provided the \SF\ is Poissonian. The key point in eliminating \SF\
in the second order moment is to expand the 2nd power in the
definition Eq.(5) of $C^{(\rm dyn)}_2(\bar {\pt})$, 
\bw \beq C^{(\rm dyn)}_2(\bar {\pt}) = \f\la \f(\sum_{m=1}^M
(\pt)_mp_m\g)^2 \g\ra = \f\la \sum_{m=1}^M (\pt)_m^2p_m^2 \g\ra 
+ \f\la \sum_{m\neq m'}^M (\pt)_m(\pt)_{m'}p_mp_{m'} \g\ra . \eeq 
Using the formulae
\beq \f\la \sum_{m=1}^M f_m p_m^p\g\ra = \f\la \sum_{m=1}^M
f_m \frac{n_m(n_m-1)\cdots(n_m-p+1)}{\la n\ra^p}\g\ra,  \eeq 
which holds for Poissonian \SF, we then get $C^{(\rm dyn)}_2(\bar
{\pt})=G_2(\bar {\pt})$, where 
\beq G_2(\bar {\pt}) = \f\la \sum_{m=1}^M (\pt)_m^2
\frac{n_m(n_m-1)}{\la n\ra^2} \g\ra
+ \f\la \sum_{m\neq m'}^M (\pt)_m (\pt)_{m'}\frac{n_mn_{m'}}{\la n\ra^2} \g\ra . \eeq   \ew 
This guarantees the equality of $G$-variance $\sigma^2_{\bar\pt G}
=G_2(\bar\pt)-G_1(\bar\pt)^2$ to dynamical ones $\sigma^2_{\bar\pt
\rm dyn}=C^{(\rm dyn)}_2(\bar\pt)-C^{(\rm dyn)}_1(\bar\pt)^2.$

 \begin{figure}
 \includegraphics[width=7.2cm]{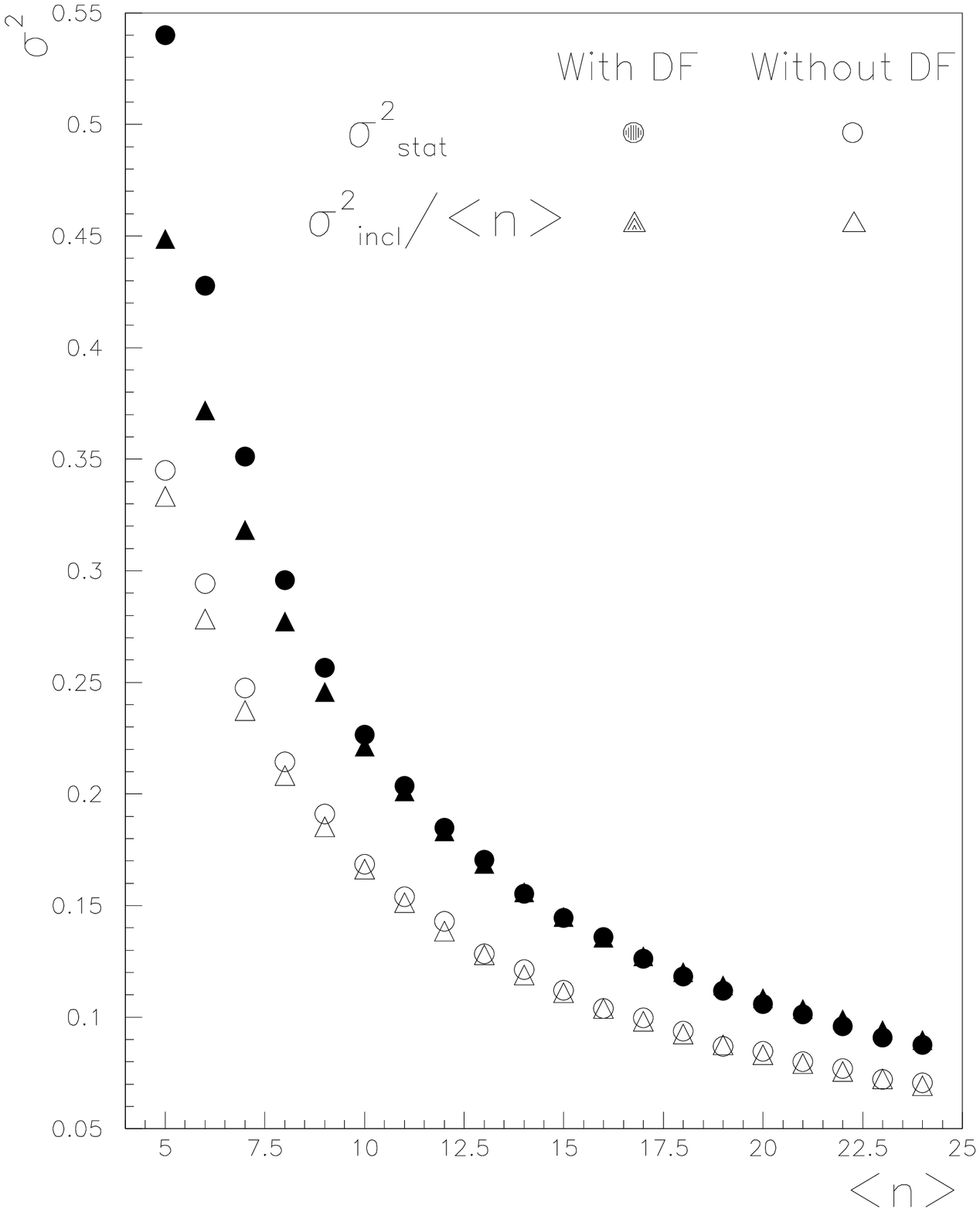}
 \caption{\label{fig:stat} Comparison of $\sigma^2_{\bar\pt \rm stat}$ and
 $\sigma^2_{\pt \rm incl}/\la n\ra$ for systems with and without \DF}
 \end{figure}

On the contrary, the \SF\ cannot be exactly gotten rid of using
the subtraction procedure. The main problem lies in the estimation
of statistical variance by Eq.(10). This equality holds only for a
pure statistical system without any dynamical fluctuation. When
dynamical fluctuations are present the $\sigma^2_{\bar{\pt}{\rm
stat}}$ and ${\sigma^2_{\pt{\rm incl}}}/{\la n\ra} $ are, in
general, unequal. For an exhibition, the $\sigma^2_{\bar{\pt}{\rm
stat}} = \sigma^2_{\bar{\pt}{\rm exp}} - \sigma^2_{\bar{\pt}{\rm
dyn}}$, {\it cf.} Fig.1, and ${\sigma^2_{\pt{\rm incl}}}/{\la
n\ra}$ are plotted in Fig.2 as solid circles and solid triangles,
respectively. They are approximately equal to each other only when
the average multiplicity is not very low, {\it e.g.} when $\la n
\ra > 12$.

In order to further clarify this point, we have made a pure
statistical model with $p_m=1/M$, {\it cf.} Eq.1. In this case
$\sigma^2_{\bar\pt \rm dyn}\equiv 0$. The corresponding
$\sigma^2_{\bar{\pt}{\rm stat}} = \sigma^2_{\bar{\pt}{\rm exp}}$
and ${\sigma^2_{\pt{\rm incl}}}/{\la n\ra}$ are plotted in Fig.2
as open circles and open triangles, respectively. The differences
between them are small up to $\la n\ra = 5$. This shows that the
discrepancy between $\sigma^2_{\bar{\pt}{\rm VKR}}$
($\sigma^2_{\bar{\pt}{\rm CERES}}$) and $\sigma^2_{\bar{\pt}{\rm
dyn}}$ is mainly due to the inappropriate estimation at low
multiplicity of $\sigma^2_{\bar{\pt}{\rm stat}}$ by
${\sigma^2_{\pt{\rm incl}}}/{\la n\ra}$ for a system containing
\DF.

 \begin{figure}
 \includegraphics[width=7.2cm]{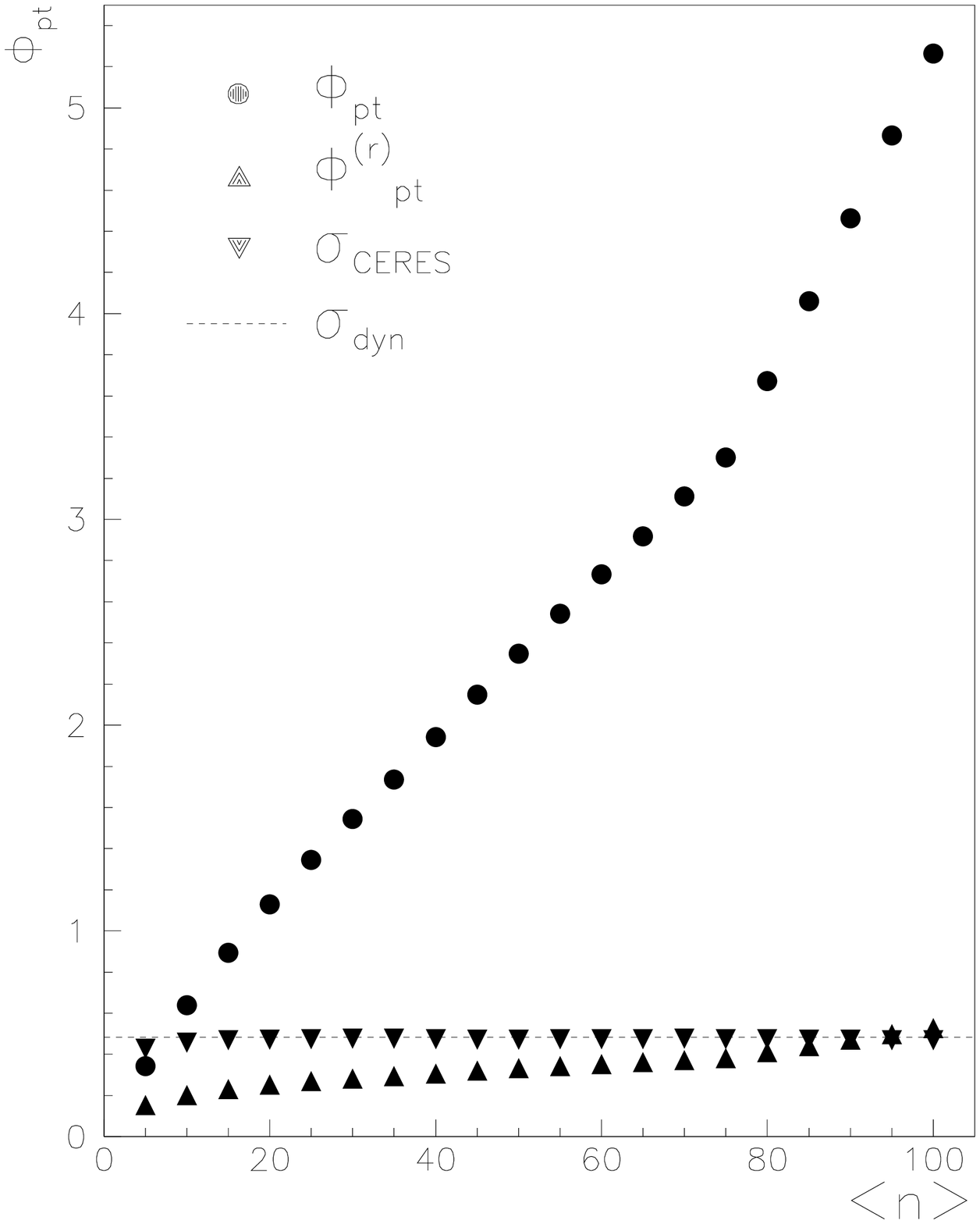}
 \caption{\label{fig:phipt} Variation of $\Phi_\pt$ and $\Phi^{\rm r}_\pt$
 with $\la n\ra$, compared with $\sigma_{_{\bar\pt \rm CERES}}$ and
 $\sigma_{\bar\pt \rm dyn}$}
 \end{figure}

Let us now turn to the discussion on the $\Phi_{\pt}$
measure~\cite{marek}\cite{NA49}. The Monte Carlo results of
$\Phi_{\pt}$ are plotted in Fig.3 as solid circles. It is strongly
multiplicity dependent, and is, therefore, not a good measure for
\DF. On accounting of Eq.(14) we divide $\Phi_{\pt}$ by the square
root of average multiplicity and define 
\beq \Phi^{\rm (r)}_{\pt} \equiv\frac{\Phi_{\pt}}{\sqrt{\la n\ra}}
\approx \sigma_{\bar{\pt}{\rm data}}-\frac{\sigma_{\pt{\rm
incl}}}{\sqrt{\la n\ra}}, \eeq 
cf. Eq.(11).

The Monte Carlo results of $\Phi^{\rm (r)}_{\pt}$ are shown in
Fig.3 as upward solid triangles together with $\sigma_{\bar{\pt}_
{\rm CERES}}= \sqrt{\sigma^2_{\bar{\pt}_ {\rm CERES}}}$ (downward
triangles) and $\sigma_{\bar{\pt} {\rm dyn}}$ (dashed line). It
can be seen from the figure that $\Phi^{\rm (r)}_{\pt}$ has a much
weaker multiplicity dependence and is closer to
$\sigma_{\bar{\pt}_{\rm dyn}}$ than $\Phi_{\pt}$. Therefore,
$\Phi^{\rm (r)}_{\pt}$ is a better measure of \DF\ than
$\Phi_{\pt}$ itself.

In this paper, various measures for the dynamical transverse
momentum fluctuation in \ebe\ analysis are compared with the
dynamical fluctuation using a Monte Carlo model.

It turns out that the $\sigma^2_{\bar{\pt} G}$ calculated from the
G-moments coincides with the dynamical variance
$\sigma^2_{\bar{\pt} \rm dyn}$, showing that the G-moment method
is effective in eliminating the statistical fluctuations coming
from the finite number of particle in a single event and thus
provides a good measure for the dynamical fluctuations.

The $\sigma^2_{\bar\pt{\rm VKR}}$ proposed by S. A. Voloshin, V.
Koch and H. G. Ritter~\cite{Voloshin} and its revised version
$\sigma^2_{\bar\pt{\rm CERES}}$~\cite{CERES} are found to be good
approximate measures for the dynamical fluctuation when the
multiplicity is not very low. Therefore, their application to
relativistic heavy ion experiments is justified.

The $\Phi_{\pt}$ divided by the square root of average
multiplicity ($\Phi^{\rm r}_{\pt}=\Phi_{\pt}/\sqrt{\la n\ra}$) is
another approximate measure for the dynamical fluctuation, which
depends on multiplicity weaker than $\Phi_{\pt}$ and is closer to
the square root of dynamical variance $\sigma_{\bar{\pt} \rm
dyn}$.

{\bf Acknowledgement} The authors thank Wu Yuanfang and Liu Feng
for helpful discussions and comments.

\ed
\begin{thebibliography}{9}
\bibitem{note2} Through out the present paper the symbol $\la x\ra$ is used to
denote the sample average of $x$, no matter $x$ is an event
variable or an inclusive variable, and the symbol $\bar x$ is
reserved for the event average of $x$.
\end{thebibliography}
